\documentclass[aps,twocolumn,showpacs,preprintnumbers,nofootinbib,float,longbibliography,superscriptaddress]{revtex4-1}

\usepackage{epsfig}
\usepackage[dvipsnames]{xcolor}
\usepackage{float,amsmath,amssymb}
\usepackage{graphicx}
\usepackage{url}
\usepackage{bbold}
\usepackage[utf8]{inputenc}
\usepackage{physics}
\newcommand{\xpom}{x_\mathbb{P}}

\newcommand{\rt}{{\mathbf{r}_\perp}}

\newcommand{\xt}{{\mathbf{x}_\perp}}

\newcommand{\yt}{{\mathbf{y}_\perp}}
\newcommand{\bt}{{\mathbf{b}_\perp}}
\newcommand{\bti}{{\mathbf{b}_{\perp,i}}}
\newcommand{\Deltat}{{\boldsymbol{\Delta}_\perp}}

\newcommand{\nc}{{N_\mathrm{c}}}
\newcommand{\jpsi}{$\mathrm{J}/\psi$ }

\usepackage[breaklinks,colorlinks,citecolor=citcolor,urlcolor=blue,linkcolor=lcolor]{hyperref}
\definecolor{lcolor}{rgb}{0.5,0,0}
\definecolor{citcolor}{rgb}{0,0.3,0.0}

\begin{document}

\title{Spatial imaging of polarized deuterons at the Electron-Ion Collider}

\author{Heikki M\"antysaari}
\affiliation{Department of Physics, University of Jyv\"askyl\"a, P.O. Box 35, 40014 University of Jyv\"askyl\"a, Finland}
\affiliation{Helsinki Institute of Physics, P.O. Box 64, 00014 University of Helsinki, Finland}

\author{Farid Salazar}
\affiliation{Institute for Nuclear Theory, University of Washington, Seattle WA 98195-1550, USA}

\author{Bj\"orn Schenke}
\affiliation{Physics Department, Brookhaven National Laboratory, Upton, NY 11973, USA}

\author{Chun Shen}
\affiliation{Department of Physics and Astronomy, Wayne State University, Detroit, Michigan 48201, USA}
\affiliation{RIKEN BNL Research Center, Brookhaven National Laboratory, Upton, NY 11973, USA}

\author{Wenbin Zhao}
\affiliation{Nuclear Science Division, Lawrence Berkeley National Laboratory, Berkeley, California 94720, USA}
\affiliation{Physics Department, University of California, Berkeley, California 94720, USA}

\begin{abstract}
We study diffractive vector meson production at small-$x$ in the collision of electrons and polarized deuterons $e+d^{\uparrow}$. We consider the polarization dependence of the nuclear wave function of the deuteron, which results in an azimuthal angular dependence of the produced vector meson when the deuteron is transversely polarized. The Fourier coefficients  extracted from the azimuthal angular dependence of the vector meson differential cross-section exhibit notable differences between longitudinally and transversely polarized deuterons. The angular dependence of the extracted effective deuteron radius provides direct insight into the structure of the polarized deuteron wave function. Furthermore, we observe slightly increased gluon saturation effects when the deuteron is longitudinally polarized compared to the transversely polarized case. The small-$x$ observables studied in this work will be accessible at the future Electron-Ion Collider.

\end{abstract}
\maketitle

\section{Introduction}
Electron-ion collisions are a powerful tool for unraveling the internal structure and dynamics of  protons and nuclei. Deep Inelastic Scattering (DIS) experiments on protons at the Hadron–Electron Ring Accelerator (HERA) at DESY in Germany have provided a detailed picture of the internal quark and gluon structure of the proton~\cite{ZEUS:2002vvv,ZEUS:2002wfj,H1:2003ksk,H1:2005dtp,H1:2009pze,H1:2013okq,H1:2015ubc,Mantysaari:2016ykx,Mantysaari:2016jaz,Schenke:2024gnj}. 
Furthermore, exclusive vector meson production in electron-nucleus collisions at high energies has been shown to offer valuable information about the geometric structure of the target nucleus across various length scales \cite{Mantysaari:2023qsq}.

The deuteron, the simplest composite nucleus consisting of a proton and a neutron, provides a convenient laboratory for investigating nucleon-nucleon forces. Its structure at large momentum fractions ($x$) has been well characterized through low-energy scattering experiments. However, our understanding of the spatial distribution of small-$x$ gluons within the deuteron remains limited.
To gain fundamental insight into the small-$x$ gluon structure of the deuteron and to furnish essential inputs for models describing deuteron-gold collisions at the Relativistic Heavy Ion Collider (RHIC) \cite{STAR:2008med,PHENIX:2017xrm,PHENIX:2018lia,Bozek:2018xzy,Zhao:2022ugy,STAR:2022pfn}, new measurements are imperative. The future Electron-Ion Collider (EIC) emerges as a promising avenue for such investigations, offering unparalleled precision in probing the gluon distribution within deuterons \cite{AbdulKhalek:2021gbh,Mantysaari:2019jhh}.

Polarization provides a unique tool to control the orientation of the deuteron. Hydrodynamic simulations of collisions involving polarized deuterons ($d^\uparrow$) and nuclei have demonstrated the valuable insights that can be gained from such setups \cite{Bozek:2018xzy,Broniowski:2019kjo}. For example, studying the elliptic flow coefficient in the azimuthal distribution of produced charged hadrons with respect to the polarization axis sheds light on the wave functions of polarized deuterons and the nature of dynamics in small systems \cite{Bozek:2018xzy,Broniowski:2019kjo}.
Electron + polarized deuteron collisions offer another rich avenue for exploration. The angular dependence of exclusive vector meson production in these collisions can effectively image the deuteron's shape in coordinate space~\cite{Zhaba:2020zaf,Gakh:2023jzo}. Moreover, such collisions offer an opportunity to probe the Generalized Parton Distributions (GPD) of the deuteron~\cite{Diehl:2003ny,Berger:2001zb,Cano:2003ju,Kirchner:2003wt,Cosyn:2018rdm}. 

Current studies at small-$x$ have focused on polarized observables in electron-proton and proton-proton collisions \cite{Kovchegov:2012ga, Brodsky:2013oya,Kovchegov:2013cva,Zhou:2013gsa,Boer:2015pni,Hatta:2016wjz,Kovchegov:2018zeq,Boussarie:2019icw,Kovchegov:2020kxg,Hagiwara:2020mqb,Kovchegov:2021iyc,Kovchegov:2022kyy,Cougoulic:2022gbk,Dumitru:2024pcv}. In this manuscript, we study electron-deuteron collisions at small-$x$ and propose a mechanism to generate azimuthal asymmetries due to the polarization-dependent nuclear wave function of the deuteron, which we take as an input for our Color Glass Condensate (CGC)~\cite{Gelis:2010nm} calculation. In the transversely polarized case and when $j_z= \pm 1$ or 0, the azimuthal anisotropies of the square of the wavefunction result in azimuthal angular asymmetries of the produced vector meson relative to the deuteron polarization axis. We compute the squared momentum transfer $t$-differential coherent and incoherent cross-sections for the production of \jpsi and $\rho$ mesons at various azimuthal angles, demonstrating the sensitivity of the observables to the spatial structure of the wave function squared. We further compute the ratio of the coherent photoproduction cross sections of $\rho$ mesons between collisions with longitudinally and transversely polarized deuteron targets. The greater extent of the longitudinally polarized deuteron along the beam direction leads to a weak saturation driven suppression of the cross section, relative to the transversely polarized case. 

This paper is organized as follows. In Sec.\,\ref{sec:vectorproduction} we review the formalism for exclusive vector meson production (coherent and incoherent) in the Color Glass Condensate. We implement the polarization-dependent nuclear wave function of the probed deuteron into the initial conditions for the color-charged correlators from which the Wilson line correlators are computed. We present the differential (both in $t$ and azimuthal angle) cross-section of \jpsi and $\rho$ vector meson production in Sec.~\ref{sec:results}. Our conclusions and outlook are presented in Sec.~\ref{sec:summary}. 

\begin{figure*}
      \includegraphics[width=0.9\columnwidth]{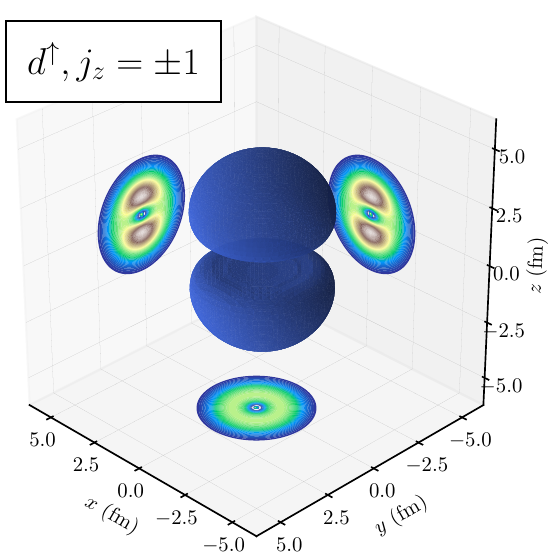}
      \includegraphics[width=0.9\columnwidth]{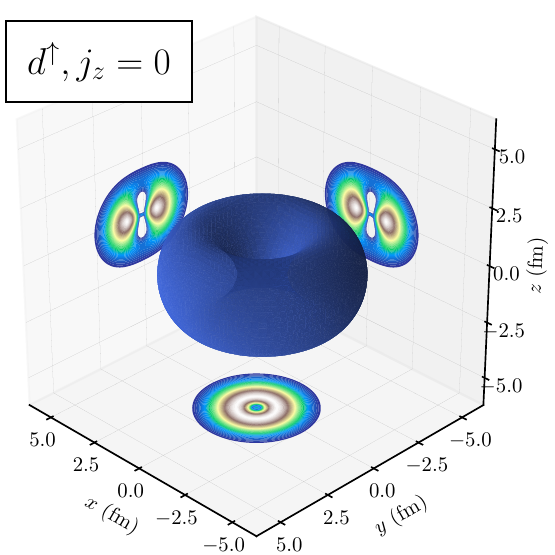}
    \caption{The nucleon density of the polarized deuteron with $j_z=\pm1$ (left) and $j_z = 0$ (right). } 
    \label{fig:edprofile}
\end{figure*}

\section{Vector meson production at high energy}
\label{sec:vectorproduction}

The differential cross-section for exclusive vector meson production in virtual photon-nucleus collisions gives us access to the spatial distribution of small-$x$ gluons inside the colliding nucleus, as well as its fluctuations. The coherent cross-section, corresponding to the process where the target remains in the same quantum state, can be obtained by averaging over the target color charge configurations $\Omega$ at the amplitude level~\cite{Good:1960ba}:

\begin{equation}
\label{eq:coherent}
     \frac{\dd \sigma^{\gamma^* + A \to V + A}}{\dd |t|}  = \frac{1}{16\pi} \left|\left\langle \mathcal{A} \right\rangle_\Omega\right|^2.
\end{equation}

The incoherent vector meson production cross-section, for which the final state of the target is different from its initial state, is obtained by subtracting the coherent contribution from the total diffractive vector meson production cross-section~\cite{Miettinen:1978jb,Caldwell:2010zza,Mantysaari:2020axf}. 
The incoherent cross-section thus has the form of a variance
\begin{multline}
\label{eq:incoherent}
     \frac{\dd \sigma^{\gamma^* + A \to V + A^*}}{\dd |t|}  = \frac{1}{16\pi} \left[
     \left\langle \left|\mathcal{A}\right|^2\right \rangle_\Omega \right. 
     - \left. \left|\left\langle \mathcal{A}\right\rangle_\Omega\right|^2 \right]\,.
\end{multline}

In the dipole picture, the production of a vector meson can be seen as the splitting of the virtual photon into a quark anti-quark pair, which then eikonally scatters off the small-$x$ gluon field of the nucleus, and finally recombines into the observed vector meson. The amplitude for this process can be written as \cite{Kowalski:2006hc,Hatta:2017cte} (see also Refs.~\cite{Mantysaari:2022kdm,Mantysaari:2021ryb} for recent developments towards NLO accuracy)
\begin{multline}
\label{eq:jpsi_amp}
    \mathcal{A} = 2i\int \dd[2]{\rt} \dd[2]{\bt}  \frac{\dd{z}}{4\pi} e^{-i \left[\bt - \left(\frac{1}{2}-z\right)\rt\right]\cdot \Deltat} \\
    \times [\Psi_V^* \Psi_\gamma](Q^2,\rt,z) N_\Omega(\rt,\bt,\xpom).
\end{multline}
Here $\rt = \xt-\yt$ is the relative vector between the quark and anti-quark positions $\xt$ and $\yt$,  $\bt = \frac{1}{2}(\xt + \yt)$ is the impact parameter vector, and $Q^2=-q^2$ is the photon virtuality. The fraction of the large photon momentum carried by the quark is given by $z$, $\xpom$ is the fraction of the target longitudinal momentum transferred to the meson in the frame where the target has a large momentum, and $\Deltat$ is the transverse momentum transfer, with $-t \approx {\boldsymbol \Delta}_\perp^2$. The $\gamma^* \to q\bar q$ splitting is described by the virtual photon light front wave function $\Psi_\gamma$~\cite{Kovchegov:2012mbw}. The vector meson wave function $\Psi_V$ is non-perturbative and needs to be modeled, introducing some uncertainty. Here, we use the Boosted Gaussian parametrization from~\cite{Kowalski:2006hc}, where the model parameters are constrained by the leptonic decay width data. 

In the Color Glass Condensate effective theory (for a review see \cite{Iancu:2003xm,Gelis:2010nm,Kovchegov:2012mbw,Morreale:2021pnn}), the dipole amplitude
\begin{align}
    &N_{\Omega}(\rt,\bt,\xpom) = \notag\\ & ~~~~~~~1 - \frac{1}{\nc} \tr \left[ V\left(\bt + \frac{\rt}{2}\right) V^\dagger\left(\bt - \frac{\rt}{2}\right) \right]\,,
\end{align}
describes the eikonal scattering of the quark anti-quark pair with the nucleus; thus, it depends on the small-$x$ gluon content of the nucleus.
In the eikonal approximation, the quark transverse position is unchanged by the scattering, and its color rotates according to the light-like Wilson line $V(\xt)$:
\begin{equation}
  V(\xt) = \mathrm{P}_{-}\left\{ \exp\left({ig\int_{-\infty}^\infty \dd{z^{-}} A^+(z^-,\xt) }\right) \right\}\,,
  \label{eq:wline_regulated}
\end{equation}
where $A^+$ is the gauge background field for the small-$x$ gluons, and $\mathrm{P}_{-}$ represents path ordering operator along the $z^-$ direction. It is obtained by solving the Yang-Mills equations where the color current is generated by large-$x$ partons, which in the CGC are treated as classical color charge density sources $\rho^a$. In covariant gauge $\partial_{\mu} A^\mu$=0, the solution is given by the Poisson equation:
\begin{align}
    A^+(x^-,\xt) = -\frac{\rho^a(x^-,\xt) t^a}{\boldsymbol{\nabla}^2 - m^2}\,,
\end{align}
where we introduced the infrared regulator $m$ to avoid the emergence of unphysical Coulomb tails. 

The color sources are obtained following the the IP-Glasma initial state description~\cite{Schenke:2012wb} used e.g.~in Refs.~\cite{Mantysaari:2022sux,Mantysaari:2023xcu,Mantysaari:2020lhf,Mantysaari:2019jhh,Mantysaari:2019csc,Mantysaari:2018zdd,Mantysaari:2016jaz,Mantysaari:2016ykx}. 
They are computed by first relating the average square color charge density to the local ($\xpom$ dependent) saturation scale extracted from the IPSat dipole-proton amplitude~\cite{Schenke:2012hg}. The proton and the neutron consist of hot spots, and the hot spot positions $\bti$ are sampled from a two-dimensional Gaussian distribution with the  width $B_{qc}$, and the center-of-mass is shifted to the origin at the end. 
In this work, we use the Maximum a Posteriori (MAP) parameter set from a Bayesian analysis where the nucleon geometry parameters at $\xpom\approx 0.0017$ are constrained by the exclusive \jpsi production data from HERA~\cite{Mantysaari:2022ffw}.

The nucleon positions are sampled according to the polarization-dependent squared wave function of the deuteron:
\begin{align}
&| \Psi(r,\theta,\phi;J_z = \pm 1)|^2 = \frac{1}{16\pi}  \left [ 4 U(r)^2  \right . \nonumber \\
&\left. ~~~~~~~~~~~~~~~~~~~~~~~~~~~~~~ - 2 \sqrt{2} \left(1-3 \cos ^2(\theta ) \right) U(r) V(r) \right. \notag \\ & \left. ~~~~~~~~~~~~~~~~~~~~~~~~~~~~~~  +\left(5-3 \cos ^2(\theta )\right) V(r)^2 \right ],  \label{eq:wv1}\\
& | \Psi(r,\theta,\phi;J_z = 0)|^2 = \frac{1}{8\pi}  \left [ 2 U(r)^2  \right . \nonumber \\ 
& ~\left . ~~~~~~~~~~~~~~~~~~~~~~~~~~~~~ +2 \sqrt{2} \left(1-3 \cos ^2(\theta )\right) U(r) V(r)\right. \notag \\ & \left. ~~~~~~~~~~~~~~~~~~~~~~~~~~~~~~+\left(1+3 \cos ^2(\theta )\right) V(r)^2 \right ],  \label{eq:wv2}
\end{align}
where $U(r)$ and $V(r)$ are the $S$-wave and $D$-wave radial functions. Following Refs.\,\cite{Bozek:2018xzy,Broniowski:2019kjo}, we use the parametrizations for $U(r)$ and $V(r)$ obtained from the Reid93 nucleon-nucleon potential listed in the literature~\cite{Zhaba:2015yxq}.
In this parametrization, the weight of the $D$-wave part in the probability distribution is around 6\%, clearly exhibiting the strong $S$-wave dominance. Figure \ref{fig:edprofile} shows the nucleon density distributions of the polarized deuteron with the polarization direction being the $z-$axis, for both $j_z = \pm1$ and $j_z = 0$. In the three planes we show 2D slices of the density for the third coordinate outside the plane set to zero, for example, the $x-y$ plane shows the density distribution at $z=0$.

This azimuthal asymmetry in the nucleon positions with respect to the polarization vector results in anisotropy in the azimuthal angle of the produced vector meson. From Eqs.\,\eqref{eq:wv1} and \eqref{eq:wv2} (also see Fig.\,\ref{fig:edprofile}) we anticipate to only get modulations of the kind $\cos(n \Phi)$ with even $n$, and $\Phi$ defined as the angle between the produced vector meson and the polarization direction of the deuteron. When identifying the square of the wave function with the probability of sampling the nucleon positions we implicitly have assumed that the target does not change its polarization during the scattering process. We left for future work studying the polarization changing case.

 In this work the longitudinal and transverse polarizations are defined in the $\gamma^* d$ center-of-momentum frame.
In the photoproduction ($Q^2=0$) limit this corresponds to the polarization states defined with respect to the electron beam as used in experiments, and as such we mostly focus on photoproduction in this work. 
When $Q^2>0$, the photon momentum is not parallel to the electron, thus a Lorentz transformation is required from the lab frame to this frame, which would mix the polarization states. 

\begin{figure*}
      \includegraphics[width=2.1\columnwidth]{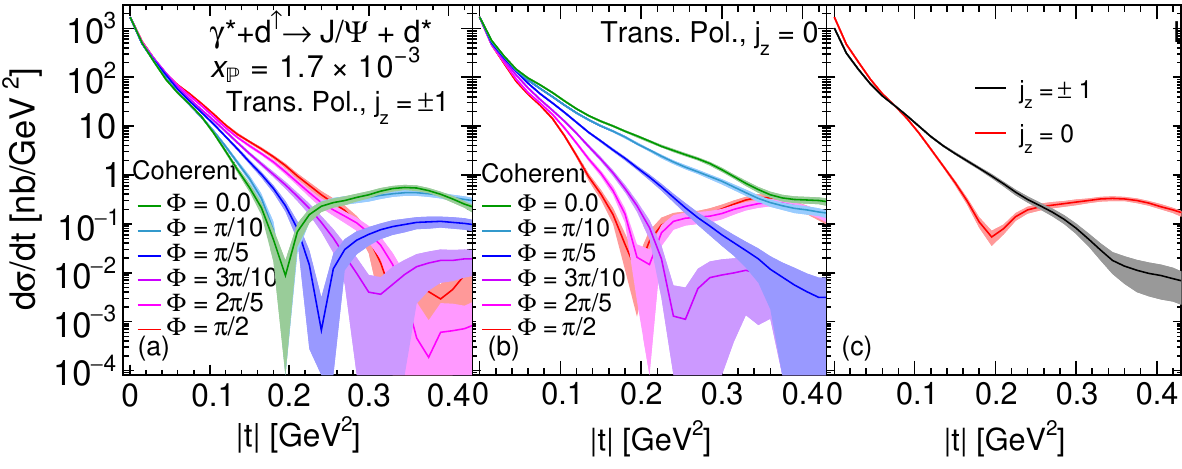}
    \caption{The angular dependence $\Phi$ of  $|t|-$spectra  of \jpsi productions for the transverse polarized deuteron with $j_z=0$ and $j_z = \pm1$. $\Phi$ is defined relative to the polarization direction of transverse polarization. The band shows the statistical uncertainty of the calculation. } 
    \label{fig:tspectra}
\end{figure*}

\section{RESULTS}
\label{sec:results}
\subsection{Imaging polarized deuterons with exclusive vector mesons}
\begin{figure}
    \centering
    \includegraphics[width=\columnwidth]{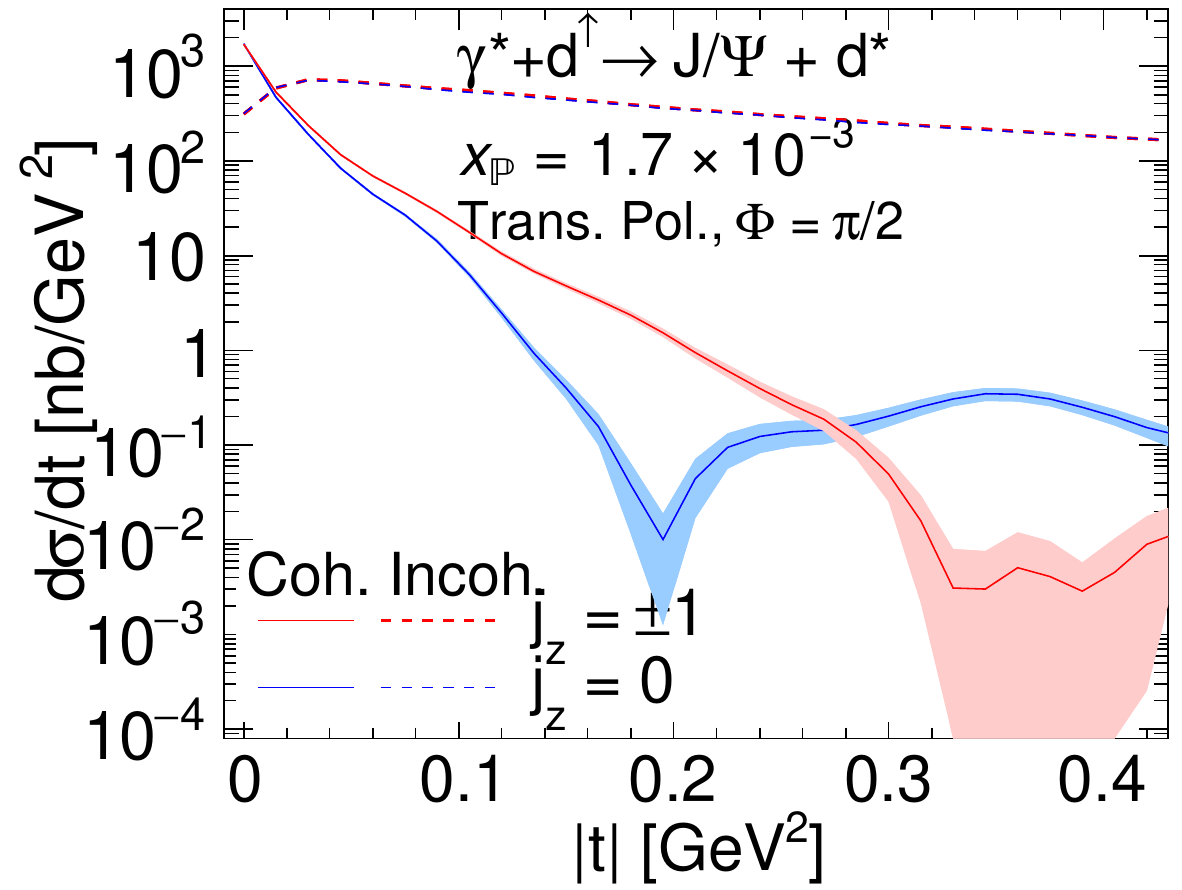}
    \caption{The coherent and incoherent \jpsi photoproduction cross
sections in transverse polarized $e+d$ collisions at $\Phi=\pi/2$.}
    \label{fig:allspectra}
\end{figure}

Figures \ref{fig:tspectra} (a) and (b) show the $\vert t\vert$-spectra of coherent exclusive \jpsi production cross-sections for the transversely polarized deuterons at $Q^2=0$ GeV$^2$, with $j_z = \pm1$ and $j_z = 0$, respectively, at various \jpsi angles relative to the deuteron's polarization.
The coherent cross-sections  exhibit distinct \jpsi angle dependencies for transverse polarization. For instance, the $|t|$ distribution becomes flatter from $\Phi = 0$ to $\Phi=\pi/2$ for deuterons with $j_z = \pm1$. 
Here $\Phi$ is the azimuthal angle between the \jpsi momentum and the deuteron polarization vector.  Conversely, deuterons with $j_z = 0$ display the opposite angular dependence with the $|t|$ slope increasing with increasing $\Phi$. This disparity arises because in transverse polarization configurations, the probed deuteron sizes vary with the vector meson's angle, and the variotion depends on $j_z$. In the case of transversely polarized deuterons with $j_z = \pm1$, as illustrated in the left panel of Figure \ref{fig:edprofile}, the probed size is larger at $\Phi=0$ than at $\Phi>0$, reaching its minimum at $\Phi = \pi/2$. Similarly, for transversely polarized deuterons with $j_z = 0$, the probed size increases from $\Phi=0$ to $\Phi=\pi/2$. Conversely, for longitudinally polarized deuterons with $j_z = 0, \pm1$, the angular dependence of the size cannot be observed, as the projection onto the $x-y$ plane is circular. Furthermore, as depicted in panel (c) of Figure~\ref{fig:allspectra}, the probed size is larger for $j_z=0$ than for $j_z = \pm1$.
This angular dependence of the coherent spectrum of the vector meson in $e+d^\uparrow$ interactions can be scrutinized in forthcoming experimental measurements at the EIC.

To compare the differences between the incoherent and coherent cross-sections, Figure \ref{fig:allspectra} displays both the coherent and incoherent \jpsi  photoproduction for transversely polarized deuterons with $j_z= 0$ and $j_z=\pm1$ at $\Phi=\pi/2$. It's evident that $j_z=0$ and $j_z=\pm1$ exhibit similar incoherent cross-sections, indicating that event-by-event fluctuations are comparable for different $j_z$ values.
However, significant differences are observed in the coherent cross-sections even at low $\vert t\vert$, where the coherent process dominates. For instance, at $\vert t \vert = 0.1 \text{ GeV}^2$, the coherent cross-section with $j_z = \pm1$ is approximately twice that with $j_z=0$, and it surpasses the incoherent cross-section. This discrepancy ensures that such differences in the coherent cross-sections at different $j_z$ values can be readily discerned through future experimental measurements.

To capture the vector meson angle dependence of the coherent cross-sections at different $\vert t\vert$ values, we perform a Fourier decomposition of the cross-section, yielding coefficients $a_n$ of various orders as a function of $\vert t\vert$,  defined as
\begin{align}
\frac{{\rm d^2} \sigma^{\gamma^* + A \to V + A}}{{\rm d} \Phi {\rm d} \vert t\vert}  =& \frac{{\rm d} \sigma^{\gamma^* + A \to V + A}}{{\rm d} \vert t\vert} \nonumber \\
&\times \frac{1}{2\pi} \left( 1 + 2 \sum_{n=1}^{\infty} a_{n} (\vert t\vert )\, e^{in\Phi} \right).
\label{eq:fourier}
\end{align}

\begin{figure}
    \centering
    \includegraphics[width=\columnwidth]{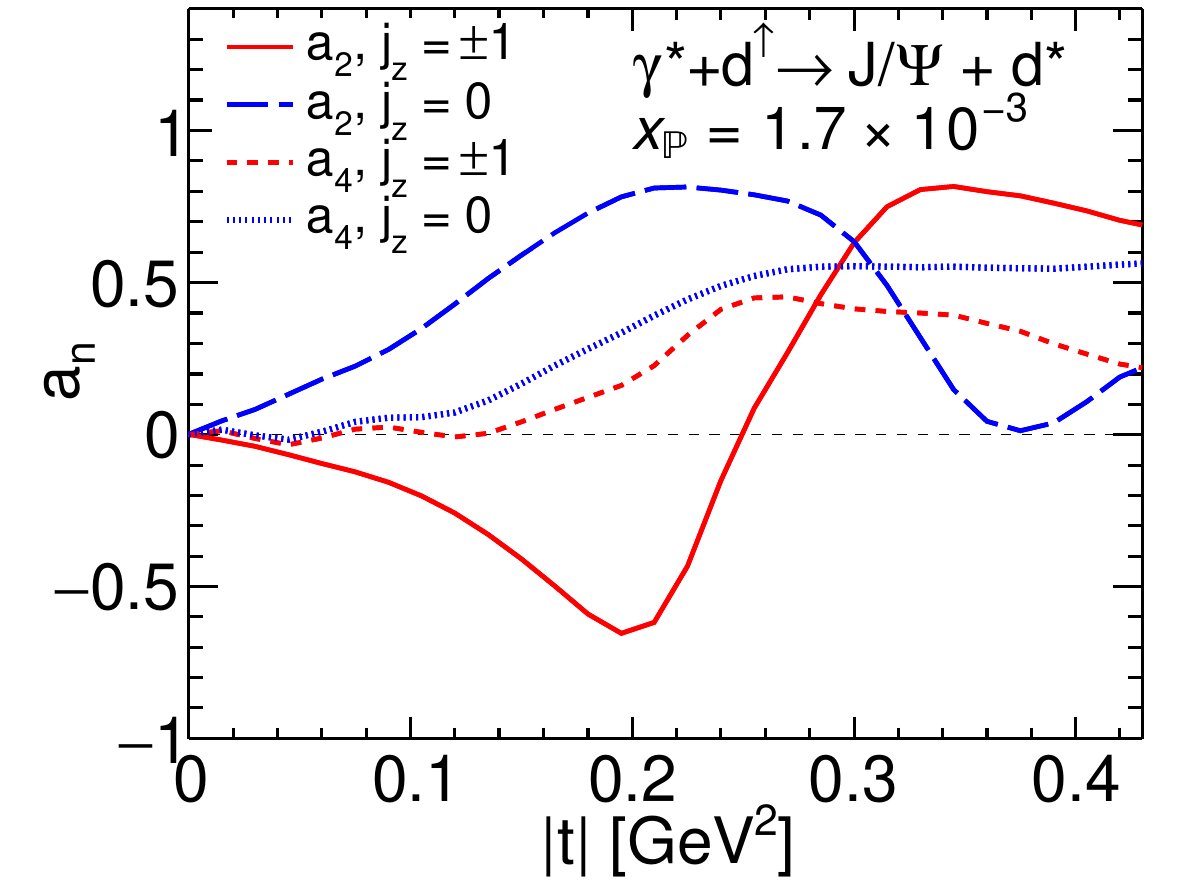}
    \caption{Real parts of the second ($a_2$) and the forth order ($a_4$) coefficients as functions of $\vert t\vert$ in transverse polarized $e+d$ collisions.}
    \label{fig:an}
\end{figure}

Figure~\ref{fig:an} shows the real parts of second and fourth order Fourier coefficients $a_{2,4}$ as a function of $\vert t \vert$ for transversely polarized deuterons with $j_z = 0$ and $\pm 1$ configurations. It is evident that at low $\vert t \vert < 0.5 \, \text{GeV}^2$, the vector meson angle dependence of the coherent cross-sections is predominantly described by the ${\rm cos(2\Phi)}$ mode. Additionally, it reveals the opposite signs of $a_2$ for $j_z = 0$ and $j_z = \pm1$, consistent with the differing angular dependence behaviors of the coherent cross-sections as depicted in Fig.\,\ref{fig:allspectra}. 
As $\vert t \vert$ increases beyond $0.4\, \text{GeV}^2$, the coherent cross-sections gradually reach their first dips, as illustrated in Fig.\,\ref{fig:allspectra}. At this stage, the ${\rm cos(4\Phi)}$ mode becomes significant, leading to large coefficients for $a_4$. In regions of large $\vert t \vert$, the $a_2$'s undergo sign changes due to the coherent cross-sections lying between the first and second dips. Specifically, $a_2$ for $j_z = \pm1$ of transverse polarization changes sign earlier than $a_2$ for $j_z = 0$. This pronounced difference in the $\vert t \vert$ dependence of Fourier coefficients between $j_z = 0$ and $\pm1$ can be examined in future measurements at the EIC.

\begin{figure}
    \centering
    \includegraphics[width=\columnwidth]{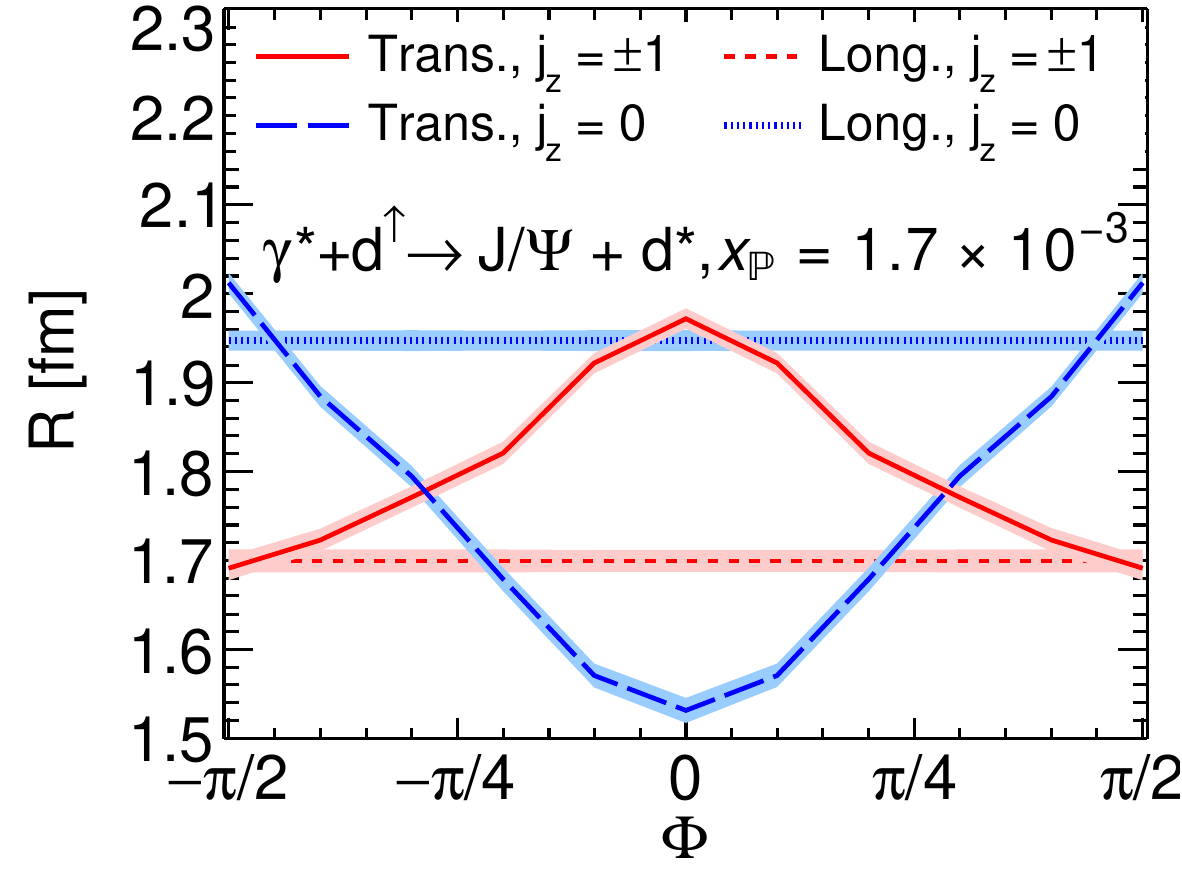}
    \caption{The effective transverse radius as a function of the $\Phi$ angle defined relative to the polarization direction of transverse polarization. }
    \label{fig:radius}
\end{figure}

To visualize the shapes of the polarized deuteron in different directions with various $j_z$ values, we extract the vector meson angular dependence of the effective radius. Following the methodology outlined in references~\cite{STAR:2017enh,STAR:2022wfe,Mantysaari:2023prg}, we obtain the effective deuteron radius by employing an empirical model to fit the calculated coherent $\vert t \vert$-spectrum. This model is characterized by the form factor of a Gaussian distribution.
We fit the coherent $t$-spectra cross sections by the formula $d\sigma/dt \sim e^{-B_{D}|t|}$  within $0 < \vert t \vert < 0.3 \, \text{GeV}^2$, and extract the effective transverse radius $R = \sqrt{2B_{D}}$. 
Although the exact geometry differs from the Gaussian distribution, this parametrization can fit the calculated spectrum well in the considered $|t|$ range.

Figure \ref{fig:radius} shows the extracted effective transverse radius $R$ as a function of $\Phi$. The extracted $R$ values for transverse polarizations exhibit a $\Phi$ dependence with a period of $\pi$. Specifically, the extracted radius for transverse polarization and $j_z = \pm1$ peaks at $\Phi = 0$ and reaches its minimum values at $\Phi = \pm\pi/2$. Conversely, the $R$ for $j_z = 0$ exhibits a dip at $\Phi = 0$ and reaches its maximum value at $\Phi = \pi/2$. In contrast, for longitudinal polarization, no $\Phi$ dependence is observed. These characteristics are discernible from the projections onto the $x-y$ and $x-z$ planes in Figure \ref{fig:edprofile}.
Experimental observation of such $\Phi$ dependence of the radius will provide direct insight into the shapes of polarized deuteron states and verify whether the spatial distribution of small $x$ gluons follows the spatial distribution of nucleons at large $x$.

\subsection{Probing saturation effects in vector meson production}
The ratio of the coherent vector meson photoproduction cross-sections at $\vert t\vert=0$ between the longitudinal and transverse polarization for polarized deuterons with $j_z=\pm1$ offers a method to probe saturation effects in $e+d^\uparrow$ collisions. Within the Color Glass Condensate framework, the coherent vector meson photoproduction cross-section at $\vert t \vert = 0$ is directly related to the total gluon content of the target as seen by the photon. Given that the effective radius (area) of longitudinal polarization with $j_z = \pm1$ is smaller than that of transverse polarization, the effective gluon density probed by the vector meson is higher in longitudinal polarization than in transverse polarization. Consequently, the gluon saturation effect is more pronounced in case of longitudinal polarization, leading to smaller vector meson production cross-sections compared to transverse polarization.

\begin{figure}
    \centering
    \includegraphics[width=\columnwidth]{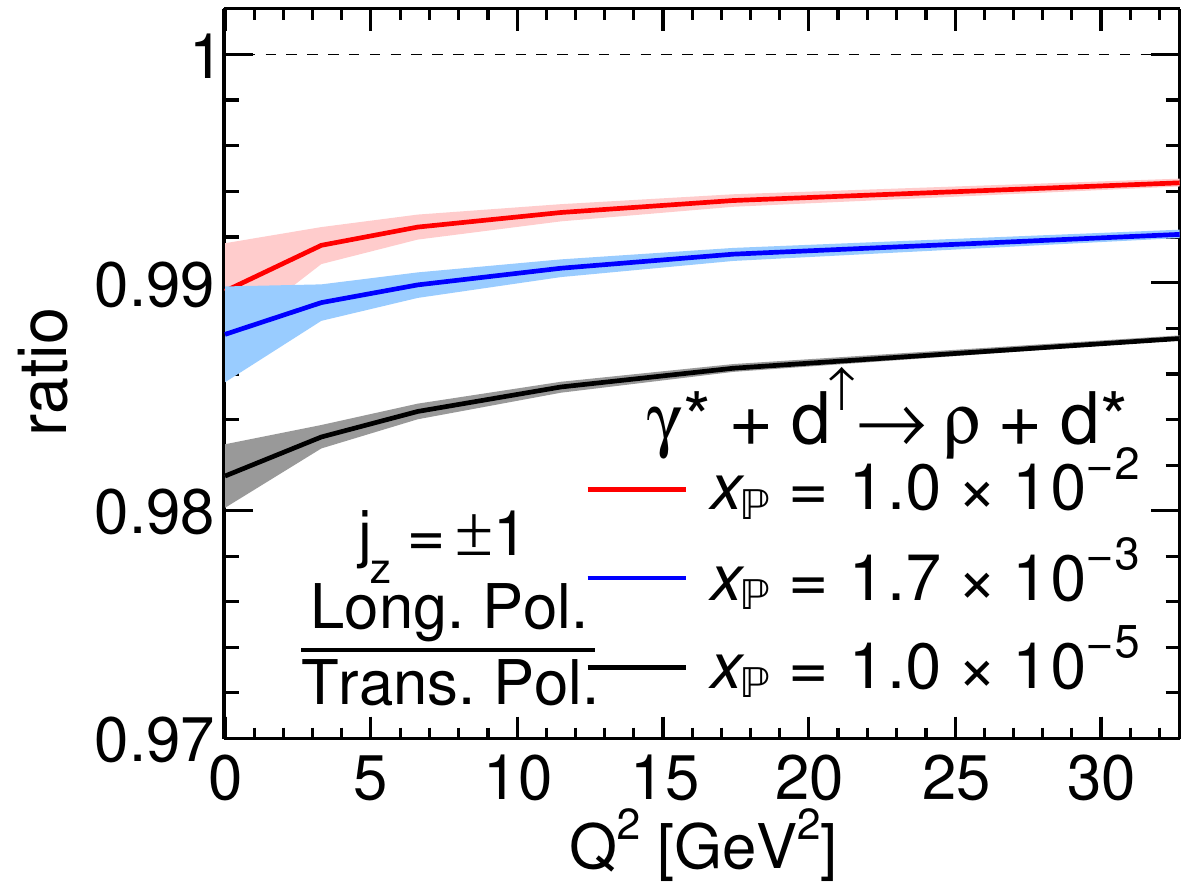}
    \caption{Longitudinal-to-transverse polarization of coherent $\rho$ photoproduction cross-section ratio at $\vert t \vert = 0$ as a function of $Q^2$ in $e+d$ collisions with $j_z=\pm1$ at different initial $\xpom$ values.  }
    \label{fig:ratio}
\end{figure}

Figure \ref{fig:ratio} depicts the longitudinal-to-transverse polarization ratio of the coherent $\rho$ cross-section at $\vert t\vert = 0$ of $j_z = \pm 1$ as a function of $Q^2$ at various initial $\xpom$ values. 
Although $\rho$ production is not strictly speaking perturbative in the photoproduction region, in Fig.~\ref{fig:ratio} we show results extrapolated to $Q^2=0$, but emphasize that our calculation is valid in the $Q^2\gtrsim 1\,\mathrm{GeV}^2$ region.
Our CGC  calculations reveal that the longitudinal-to-transverse ratio is consistently below one as expected. Furthermore, the ratios exhibit a slight increasing trend as a function of $Q^2$. This trend arises because the vector meson production at a $Q^2$ probes gluons with a size of $1/Q^2$ in the target. Consequently, smaller $Q^2$ values probe larger gluons, thereby enhancing the overlap between gluons and resulting in stronger gluon saturation effects. Additionally, we observe stronger suppression in the CGC calculations for smaller $\xpom$ values. However, the magnitude of the ratio remains within 2\%, given that the deuteron comprises only two nucleons.
Note that as discussed in Sec.~\ref{sec:vectorproduction}, here the deuteron polarization is defined in the $\gamma^* d$ center-of-momentum frame.
That means that for $Q^2>0$ the polarization states in the laboratory frame and in the $\gamma^*d$ rest frame would mix, resulting in even smaller saturation effects for the cross section ratio. Thus, our result can be seen as an upper limit for the expected saturation effect at finite $Q^2$.
In the future, if larger nuclei with elongated polarization states can be employed, they may exhibit greater saturation signals. 

\section{Summary}
\label{sec:summary}
We  have explored the potential of Electron-Ion Collider (EIC) experiments to probe the spatial distribution of small-$x$ gluons within polarized deuterons through exclusive vector meson production in $e+d^{\uparrow}$ collisions. 
Our results demonstrate that the angular dependence of coherent vector meson photoproduction cross-sections in transversely polarized deuterons exhibits distinct features for $j_z = 0$ and $j_z = \pm1$ configurations. This angular dependence arises from the variation in the probed deuteron size with the vector meson's angle relative to the polarization direction. 
Specifically, the Fourier coefficients and the angular dependence of the effective radius exhibit notable differences between longitudinally and transversely polarized deuterons, suggesting that future experimental measurements at the EIC could shed light on the internal structure of the deuteron with unprecedented precision.

Furthermore, our analysis highlights the potential of vector meson production in probing saturation effects in small-$x$ gluon distributions. The longitudinal-to-transverse polarization ratio of the coherent $\rho$ cross-section at $\vert t\vert = 0$ in $j_z = \pm1$ indicates that the gluon recombination (saturation) effects are more pronounced in longitudinal polarization, leading to slightly smaller cross-sections compared to transverse polarization. This observation underscores the role polarized targets could play in understanding the dynamics of gluon saturation.

In summary, our findings highlight the potential of exclusive vector meson production to reveal the spatial distribution of small-$x$ gluons within polarized deuterons, allowing access to the spatial wave function squared of different polarization states. Corresponding measurements will be possible at the future EIC, and similar studies in the large $x$ regime also at JLab. Extending such studies to larger polarized nuclei may offer insights into saturation phenomena, providing an new handle for understanding gluon dynamics at small momentum fractions.

\begin{acknowledgments}
We thank  Y. Guo, Y. Kovchegov, S. Li, L. Wang, F. Yuan, and J. Zhou for valuable discussions. We thank G. Giacalone for providing the script to plot Figure~\ref{fig:edprofile}.
This material is based upon work supported by the U.S. Department of Energy, Office of Science, Office of Nuclear Physics, under DOE Contract No.~DE-SC0012704 (B.P.S.) and Award No.~DE-SC0021969 (C.S.), and within the framework of the Saturated Glue (SURGE) Topical Theory Collaboration.
C.S. acknowledges a DOE Office of Science Early Career Award. 
H.M. is supported by the Research Council of Finland, the Centre of Excellence in Quark Matter, and projects 338263 and 346567, and under the European Union’s Horizon 2020 research and innovation programme by the European Research Council (ERC, grant agreements No. ERC-2023-101123801 GlueSatLight and ERC-2018-ADG-835105 YoctoLHC) and by the STRONG-2020 project (grant agreement No 824093) and wishes to thank the EIC Theory Institute at BNL for its hospitality during the early stages of this work.
F.S. is supported by the Institute for Nuclear Theory’s U.S. DOE under Grant No. DE-FG02-00ER41132.
W.B.Z. is supported by DOE under Contract No. DE-AC02-05CH11231, by NSF under Grant No. OAC-2004571 within the X-SCAPE Collaboration, and within the framework of the SURGE Topical Theory Collaboration. 
The content of this article does not reflect the official opinion of the European Union and responsibility for the information and views expressed therein lies entirely with the authors.
This research was done using resources provided by the Open Science Grid (OSG)~\cite{Pordes:2007zzb, Sfiligoi:2009cct}, which is supported by the National Science Foundation award \#2030508.
\end{acknowledgments}
\bibliographystyle{JHEP-2modlong.bst}

\bibliography{refs}

@article{Bozek:2018xzy,
    author = "Bozek, Piotr and Broniowski, Wojciech",
    title = "{Elliptic Flow in Ultrarelativistic Collisions with Polarized Deuterons}",
    eprint = "1808.09840",
    archivePrefix = "arXiv",
    primaryClass = "nucl-th",
    doi = "10.1103/PhysRevLett.121.202301",
    journal = "Phys. Rev. Lett.",
    volume = "121",
    number = "20",
    pages = "202301",
    year = "2018"
}

@article{Schenke:2024gnj,
    author = {Schenke, Bj\"orn and M\"antysaari, Heikki and Salazar, Farid and Shen, Chun and Zhao, Wenbin},
    title = "{Vector meson production in ultraperipheral heavy ion collisions}",
    booktitle = "{1st International Workshop on the physics of Ultra Peripheral Collisions}",
    eprint = "2404.10833",
    archivePrefix = "arXiv",
    primaryClass = "nucl-th",
    month = "4",
    year = "2024"
}

@article{Broniowski:2019kjo,
    author = "Broniowski, Wojciech and Bo\.zek, Piotr",
    title = "{Elliptic flow in ultrarelativistic collisions with light polarized nuclei}",
    eprint = "1906.09045",
    archivePrefix = "arXiv",
    primaryClass = "nucl-th",
    doi = "10.1103/PhysRevC.101.024901",
    journal = "Phys. Rev. C",
    volume = "101",
    number = "2",
    pages = "024901",
    year = "2020"
}

@article{Mantysaari:2023qsq,
    author = {M\"antysaari, Heikki and Schenke, Bj\"orn and Shen, Chun and Zhao, Wenbin},
    title = "{Multiscale Imaging of Nuclear Deformation at the Electron-Ion Collider}",
    eprint = "2303.04866",
    archivePrefix = "arXiv",
    primaryClass = "nucl-th",
    doi = "10.1103/PhysRevLett.131.062301",
    journal = "Phys. Rev. Lett.",
    volume = "131",
    number = "6",
    pages = "062301",
    year = "2023"
}

@article{AbdulKhalek:2021gbh,
    author = "Abdul Khalek, R. and others",
    title = "{Science Requirements and Detector Concepts for the Electron-Ion Collider}: {EIC Yellow Report}",
    eprint = "2103.05419",
    archivePrefix = "arXiv",
    primaryClass = "physics.ins-det",
    reportNumber = "BNL-220990-2021-FORE, JLAB-PHY-21-3198, LA-UR-21-20953",
    doi = "10.1016/j.nuclphysa.2022.122447",
    journal = "Nucl. Phys. A",
    volume = "1026",
    pages = "122447",
    year = "2022"
}

@article{Mantysaari:2019jhh,
    author = {M\"antysaari, Heikki and Schenke, Bj\"orn},
    title = "{Accessing the gluonic structure of light nuclei at a future electron-ion collider}",
    eprint = "1910.03297",
    archivePrefix = "arXiv",
    primaryClass = "hep-ph",
    doi = "10.1103/PhysRevC.101.015203",
    journal = "Phys. Rev. C",
    volume = "101",
    number = "1",
    pages = "015203",
    year = "2020"
}

@article{Zhaba:2015yxq,
    author = "Zhaba, V. I.",
    title = "{New analytical forms of wave function in coordinate space and tensor polarization of deuteron}",
    eprint = "1512.08980",
    archivePrefix = "arXiv",
    primaryClass = "nucl-th",
    doi = "10.1142/S021773231650139X",
    journal = "Mod. Phys. Lett. A",
    volume = "31",
    number = "25",
    pages = "1650139",
    year = "2016"
}

@book{Kovchegov:2012mbw,
    author = "Kovchegov, Yuri V. and Levin, Eugene",
    title = "{Quantum Chromodynamics at High Energy}",
    doi = "10.1017/9781009291446",
    isbn = "978-1-00-929144-6, 978-1-00-929141-5, 978-1-00-929142-2, 978-0-521-11257-4, 978-1-139-55768-9",
    publisher = "Oxford University Press",
    volume = "33",
    year = "2013"
}

@article{Hatta:2017cte,
    author = "Hatta, Yoshitaka and Xiao, Bo-Wen and Yuan, Feng",
    title = "{Gluon Tomography from Deeply Virtual Compton Scattering at Small-x}",
    eprint = "1703.02085",
    archivePrefix = "arXiv",
    primaryClass = "hep-ph",
    reportNumber = "YITP-17-22",
    doi = "10.1103/PhysRevD.95.114026",
    journal = "Phys. Rev. D",
    volume = "95",
    number = "11",
    pages = "114026",
    year = "2017"
}

@article{Mantysaari:2023prg,
    author = {M\"antysaari, Heikki and Salazar, Farid and Schenke, Bj\"orn and Shen, Chun and Zhao, Wenbin},
    title = "{Effects of nuclear structure and quantum interference on diffractive vector meson production in ultraperipheral nuclear collisions}",
    eprint = "2310.15300",
    archivePrefix = "arXiv",
    primaryClass = "nucl-th",
    doi = "10.1103/PhysRevC.109.024908",
    journal = "Phys. Rev. C",
    volume = "109",
    number = "2",
    pages = "024908",
    year = "2024"
}

@article{H1:2009pze,
    author = "Aaron, F. D. and others",
    collaboration = "H1, ZEUS",
    title = "{Combined Measurement and QCD Analysis of the Inclusive $e^\pm p$ Scattering Cross Sections at HERA}",
    eprint = "0911.0884",
    archivePrefix = "arXiv",
    primaryClass = "hep-ex",
    reportNumber = "DESY-09-158",
    doi = "10.1007/JHEP01(2010)109",
    journal = "JHEP",
    volume = "01",
    pages = "109",
    year = "2010"
}

@article{H1:2015ubc,
    author = "Abramowicz, H. and others",
    collaboration = "H1, ZEUS",
    title = "{Combination of measurements of inclusive deep inelastic ${e^{\pm }p}$ scattering cross sections and QCD analysis of HERA data}",
    eprint = "1506.06042",
    archivePrefix = "arXiv",
    primaryClass = "hep-ex",
    reportNumber = "DESY-15-039",
    doi = "10.1140/epjc/s10052-015-3710-4",
    journal = "Eur. Phys. J. C",
    volume = "75",
    number = "12",
    pages = "580",
    year = "2015"
}

@article{Mantysaari:2016ykx,
    author = {M\"antysaari, Heikki and Schenke, Bj\"orn},
    title = "{Evidence of strong proton shape fluctuations from incoherent diffraction}",
    eprint = "1603.04349",
    archivePrefix = "arXiv",
    primaryClass = "hep-ph",
    doi = "10.1103/PhysRevLett.117.052301",
    journal = "Phys. Rev. Lett.",
    volume = "117",
    number = "5",
    pages = "052301",
    year = "2016"
}

@article{Mantysaari:2016jaz,
    author = {M\"antysaari, Heikki and Schenke, Bj\"orn},
    title = "{Revealing proton shape fluctuations with incoherent diffraction at high energy}",
    eprint = "1607.01711",
    archivePrefix = "arXiv",
    primaryClass = "hep-ph",
    doi = "10.1103/PhysRevD.94.034042",
    journal = "Phys. Rev. D",
    volume = "94",
    number = "3",
    pages = "034042",
    year = "2016"
}

@article{Gelis:2010nm,
    author = "Gelis, Francois and Iancu, Edmond and Jalilian-Marian, Jamal and Venugopalan, Raju",
    title = "{The Color Glass Condensate}",
    eprint = "1002.0333",
    archivePrefix = "arXiv",
    primaryClass = "hep-ph",
    doi = "10.1146/annurev.nucl.010909.083629",
    journal = "Ann. Rev. Nucl. Part. Sci.",
    volume = "60",
    pages = "463--489",
    year = "2010"
}

@article{STAR:2008med,
    author = "Abelev, B. I. and others",
    collaboration = "STAR",
    title = "{Systematic Measurements of Identified Particle Spectra in $p p, d^+$ Au and Au+Au Collisions from STAR}",
    eprint = "0808.2041",
    archivePrefix = "arXiv",
    primaryClass = "nucl-ex",
    doi = "10.1103/PhysRevC.79.034909",
    journal = "Phys. Rev. C",
    volume = "79",
    pages = "034909",
    year = "2009"
}

@article{PHENIX:2017xrm,
    author = "Aidala, C. and others",
    collaboration = "PHENIX",
    title = "{Measurements of Multiparticle Correlations in $d+\mathrm{Au}$ Collisions at 200, 62.4, 39, and 19.6 GeV and $p+\mathrm{Au}$ Collisions at 200 GeV and Implications for Collective Behavior}",
    eprint = "1707.06108",
    archivePrefix = "arXiv",
    primaryClass = "nucl-ex",
    doi = "10.1103/PhysRevLett.120.062302",
    journal = "Phys. Rev. Lett.",
    volume = "120",
    number = "6",
    pages = "062302",
    year = "2018"
}

@article{PHENIX:2018lia,
    author = "Aidala, C. and others",
    collaboration = "PHENIX",
    title = "{Creation of quark\textendash{}gluon plasma droplets with three distinct geometries}",
    eprint = "1805.02973",
    archivePrefix = "arXiv",
    primaryClass = "nucl-ex",
    doi = "10.1038/s41567-018-0360-0",
    journal = "Nature Phys.",
    volume = "15",
    number = "3",
    pages = "214--220",
    year = "2019"
}

@article{Zhao:2022ugy,
    author = {Zhao, Wenbin and Ryu, Sangwook and Shen, Chun and Schenke, Bj\"orn},
    title = "{3D structure of anisotropic flow in small collision systems at energies available at the BNL Relativistic Heavy Ion Collider}",
    eprint = "2211.16376",
    archivePrefix = "arXiv",
    primaryClass = "nucl-th",
    doi = "10.1103/PhysRevC.107.014904",
    journal = "Phys. Rev. C",
    volume = "107",
    number = "1",
    pages = "014904",
    year = "2023"
}

@article{STAR:2022pfn,
    author = "Abdulhamid, M. I. and others",
    collaboration = "STAR",
    title = "{Measurements of the Elliptic and Triangular Azimuthal Anisotropies in Central ${}^3$He+Au, d+Au and p+Au Collisions at $\sqrt{s_{NN}}=200\,\,GeV$}",
    eprint = "2210.11352",
    archivePrefix = "arXiv",
    primaryClass = "nucl-ex",
    doi = "10.1103/PhysRevLett.130.242301",
    journal = "Phys. Rev. Lett.",
    volume = "130",
    number = "24",
    pages = "242301",
    year = "2023"
}

@article{Zhaba:2020zaf,
    author = "Zhaba, V. I.",
    title = "{Born Values for Vector and Tensor Asymmetries in Electron-deuteron Scattering}",
    journal = "Prob. Atomic Sci. Technol.",
    volume = "2020",
    number = "5",
    pages = "19--22",
    year = "2020"
}

@article{Gakh:2023jzo,
    author = "Gakh, G. I. and Konchatnij, M. I. and Merenkov, N. P. and Tomasi-Gustafsson, E. and Gakh, A. G.",
    title = "{Polarization effects in elastic deuteron-electron scattering}",
    eprint = "2311.01102",
    archivePrefix = "arXiv",
    primaryClass = "nucl-th",
    doi = "10.1103/PhysRevC.109.065203",
    journal = "Phys. Rev. C",
    volume = "109",
    number = "6",
    pages = "065203",
    year = "2024"
}

@article{Diehl:2003ny,
    author = "Diehl, M.",
    title = "{Generalized parton distributions}",
    eprint = "hep-ph/0307382",
    archivePrefix = "arXiv",
    reportNumber = "DESY-THESIS-2003-018",
    doi = "10.1016/j.physrep.2003.08.002",
    journal = "Phys. Rept.",
    volume = "388",
    pages = "41--277",
    year = "2003"
}

@article{Good:1960ba,
    author = "Good, M. L. and Walker, W. D.",
    title = "{Diffraction disssociation of beam particles}",
    doi = "10.1103/PhysRev.120.1857",
    journal = "Phys. Rev.",
    volume = "120",
    pages = "1857--1860",
    year = "1960"
}

@article{Miettinen:1978jb,
    author = "Miettinen, Hannu I. and Pumplin, Jon",
    title = "{Diffraction Scattering and the Parton Structure of Hadrons}",
    reportNumber = "FERMILAB-PUB-78-021-T",
    doi = "10.1103/PhysRevD.18.1696",
    journal = "Phys. Rev. D",
    volume = "18",
    pages = "1696",
    year = "1978"
}

@article{Caldwell:2010zza,
    author = "Caldwell, A. and Kowalski, H.",
    title = "{Investigating the gluonic structure of nuclei via $\mathrm{J}/\psi$ scattering}",
    doi = "10.1103/PhysRevC.81.025203",
    journal = "Phys. Rev. C",
    volume = "81",
    pages = "025203",
    year = "2010"
}

@article{Mantysaari:2020axf,
    author = {M\"antysaari, Heikki},
    title = "{Review of proton and nuclear shape fluctuations at high energy}",
    eprint = "2001.10705",
    archivePrefix = "arXiv",
    primaryClass = "hep-ph",
    doi = "10.1088/1361-6633/aba347",
    journal = "Rept. Prog. Phys.",
    volume = "83",
    number = "8",
    pages = "082201",
    year = "2020"
}

@article{Mantysaari:2022kdm,
    author = {M\"antysaari, Heikki and Penttala, Jani},
    title = "{Complete calculation of exclusive heavy vector meson production at next-to-leading order in the dipole picture}",
    eprint = "2204.14031",
    archivePrefix = "arXiv",
    primaryClass = "hep-ph",
    doi = "10.1007/JHEP08(2022)247",
    journal = "JHEP",
    volume = "08",
    pages = "247",
    year = "2022"
}

@article{Mantysaari:2021ryb,
    author = {M\"antysaari, Heikki and Penttala, Jani},
    title = "{Exclusive heavy vector meson production at next-to-leading order in the dipole picture}",
    eprint = "2104.02349",
    archivePrefix = "arXiv",
    primaryClass = "hep-ph",
    doi = "10.1016/j.physletb.2021.136723",
    journal = "Phys. Lett. B",
    volume = "823",
    pages = "136723",
    year = "2021"
}

@article{Kowalski:2006hc,
    author = "Kowalski, H. and Motyka, L. and Watt, G.",
    title = "{Exclusive diffractive processes at HERA within the dipole picture}",
    eprint = "hep-ph/0606272",
    archivePrefix = "arXiv",
    reportNumber = "DESY-06-095",
    doi = "10.1103/PhysRevD.74.074016",
    journal = "Phys. Rev. D",
    volume = "74",
    pages = "074016",
    year = "2006"
}

@article{Schenke:2012hg,
    author = "Schenke, Bjoern and Tribedy, Prithwish and Venugopalan, Raju",
    title = "{Event-by-event gluon multiplicity, energy density, and eccentricities in ultrarelativistic heavy-ion collisions}",
    eprint = "1206.6805",
    archivePrefix = "arXiv",
    primaryClass = "hep-ph",
    doi = "10.1103/PhysRevC.86.034908",
    journal = "Phys. Rev. C",
    volume = "86",
    pages = "034908",
    year = "2012"
}

@article{Schenke:2012wb,
    author = "Schenke, Bjoern and Tribedy, Prithwish and Venugopalan, Raju",
    title = "{Fluctuating Glasma initial conditions and flow in heavy ion collisions}",
    eprint = "1202.6646",
    archivePrefix = "arXiv",
    primaryClass = "nucl-th",
    doi = "10.1103/PhysRevLett.108.252301",
    journal = "Phys. Rev. Lett.",
    volume = "108",
    pages = "252301",
    year = "2012"
}

@article{Mantysaari:2022sux,
    author = {M\"antysaari, Heikki and Salazar, Farid and Schenke, Bj\"orn},
    title = "{Nuclear geometry at high energy from exclusive vector meson production}",
    eprint = "2207.03712",
    archivePrefix = "arXiv",
    primaryClass = "hep-ph",
    doi = "10.1103/PhysRevD.106.074019",
    journal = "Phys. Rev. D",
    volume = "106",
    number = "7",
    pages = "074019",
    year = "2022"
}

@article{Mantysaari:2020lhf,
    author = {M\"antysaari, Heikki and Roy, Kaushik and Salazar, Farid and Schenke, Bj\"orn},
    title = "{Gluon imaging using azimuthal correlations in diffractive scattering at the Electron-Ion Collider}",
    eprint = "2011.02464",
    archivePrefix = "arXiv",
    primaryClass = "hep-ph",
    doi = "10.1103/PhysRevD.103.094026",
    journal = "Phys. Rev. D",
    volume = "103",
    number = "9",
    pages = "094026",
    year = "2021"
}

@article{Mantysaari:2019csc,
    author = {M\"antysaari, Heikki and Mueller, Niklas and Schenke, Bj\"orn},
    title = "{Diffractive Dijet Production and Wigner Distributions from the Color Glass Condensate}",
    eprint = "1902.05087",
    archivePrefix = "arXiv",
    primaryClass = "hep-ph",
    doi = "10.1103/PhysRevD.99.074004",
    journal = "Phys. Rev. D",
    volume = "99",
    number = "7",
    pages = "074004",
    year = "2019"
}

@article{Mantysaari:2018zdd,
    author = {M\"antysaari, Heikki and Schenke, Bj\"orn},
    title = "{Confronting impact parameter dependent JIMWLK evolution with HERA data}",
    eprint = "1806.06783",
    archivePrefix = "arXiv",
    primaryClass = "hep-ph",
    doi = "10.1103/PhysRevD.98.034013",
    journal = "Phys. Rev. D",
    volume = "98",
    number = "3",
    pages = "034013",
    year = "2018"
}

@article{Mantysaari:2022ffw,
    author = {M\"antysaari, Heikki and Schenke, Bj\"orn and Shen, Chun and Zhao, Wenbin},
    title = "{Bayesian inference of the fluctuating proton shape}",
    eprint = "2202.01998",
    archivePrefix = "arXiv",
    primaryClass = "hep-ph",
    doi = "10.1016/j.physletb.2022.137348",
    journal = "Phys. Lett. B",
    volume = "833",
    pages = "137348",
    year = "2022"
}

@article{STAR:2017enh,
    author = "Adamczyk, L. and others",
    collaboration = "STAR",
    title = "{Coherent diffractive photoproduction of $\rho^{0}$mesons on gold nuclei at 200 GeV/nucleon-pair at the Relativistic Heavy Ion Collider}",
    eprint = "1702.07705",
    archivePrefix = "arXiv",
    primaryClass = "nucl-ex",
    doi = "10.1103/PhysRevC.96.054904",
    journal = "Phys. Rev. C",
    volume = "96",
    number = "5",
    pages = "054904",
    year = "2017"
}

@article{STAR:2022wfe,
    author = "Abdallah, Mohamed and others",
    collaboration = "STAR",
    title = "{Tomography of ultrarelativistic nuclei with polarized photon-gluon collisions}",
    eprint = "2204.01625",
    archivePrefix = "arXiv",
    primaryClass = "nucl-ex",
    doi = "10.1126/sciadv.abq3903",
    journal = "Sci. Adv.",
    volume = "9",
    number = "1",
    pages = "eabq3903",
    year = "2023"
}

@article{Pordes:2007zzb,
    author = "Pordes, Ruth and others",
    editor = "Keyes, David E.",
    title = "{The Open Science Grid}",
    reportNumber = "FERMILAB-CONF-07-217-CD",
    doi = "10.1088/1742-6596/78/1/012057",
    journal = "J. Phys. Conf. Ser.",
    volume = "78",
    pages = "012057",
    year = "2007"
}

@article{H1:2005dtp,
    author = "Aktas, A. and others",
    collaboration = "H1",
    title = "{Elastic $\mathrm{J}/\psi$ production at HERA}",
    eprint = "hep-ex/0510016",
    archivePrefix = "arXiv",
    reportNumber = "DESY-05-161",
    doi = "10.1140/epjc/s2006-02519-5",
    journal = "Eur. Phys. J. C",
    volume = "46",
    pages = "585--603",
    year = "2006"
}

@article{ZEUS:2002wfj,
    author = "Chekanov, S. and others",
    collaboration = "ZEUS",
    title = "{Exclusive photoproduction of $\mathrm{J}/\psi$ mesons at HERA}",
    eprint = "hep-ex/0201043",
    archivePrefix = "arXiv",
    reportNumber = "DESY-02-008",
    doi = "10.1007/s10052-002-0953-7",
    journal = "Eur. Phys. J. C",
    volume = "24",
    pages = "345--360",
    year = "2002"
}

@article{ZEUS:2002vvv,
    author = "Chekanov, S. and others",
    collaboration = "ZEUS",
    title = "{Measurement of proton dissociative diffractive photoproduction of vector mesons at large momentum transfer at HERA}",
    eprint = "hep-ex/0205081",
    archivePrefix = "arXiv",
    reportNumber = "DESY-02-072",
    doi = "10.1140/epjc/s2002-01079-0",
    journal = "Eur. Phys. J. C",
    volume = "26",
    pages = "389--409",
    year = "2003"
}

@article{H1:2003ksk,
    author = "Aktas, A. and others",
    collaboration = "H1",
    title = "{Diffractive photoproduction of $J/\psi$ mesons with large momentum transfer at HERA}",
    eprint = "hep-ex/0306013",
    archivePrefix = "arXiv",
    reportNumber = "DESY-03-061",
    doi = "10.1016/j.physletb.2003.06.056",
    journal = "Phys. Lett. B",
    volume = "568",
    pages = "205--218",
    year = "2003"
}

@article{H1:2013okq,
    author = "Alexa, C. and others",
    collaboration = "H1",
    title = "{Elastic and Proton-Dissociative Photoproduction of J/psi Mesons at HERA}",
    eprint = "1304.5162",
    archivePrefix = "arXiv",
    primaryClass = "hep-ex",
    reportNumber = "DESY-13-058",
    doi = "10.1140/epjc/s10052-013-2466-y",
    journal = "Eur. Phys. J. C",
    volume = "73",
    number = "6",
    pages = "2466",
    year = "2013"
}

@article{Sfiligoi:2009cct,
    author = "Sfiligoi, Igor and Bradley, Daniel C. and Holzman, Burt and Mhashilkar, Parag and Padhi, Sanjay and Wurthwrin, Frank",
    title = "{The pilot way to Grid resources using glideinWMS}",
    reportNumber = "FERMILAB-CONF-09-373-CD",
    doi = "10.1109/CSIE.2009.950",
    journal = "WRI World Congress",
    volume = "2",
    pages = "428--432",
    year = "2009"
}

@article{Berger:2001zb,
    author = "Berger, Edgar R. and Cano, F. and Diehl, M. and Pire, B.",
    title = "{Generalized parton distributions in the deuteron}",
    eprint = "hep-ph/0106192",
    archivePrefix = "arXiv",
    reportNumber = "CPHT-S029-0601, DESY-01-082",
    doi = "10.1103/PhysRevLett.87.142302",
    journal = "Phys. Rev. Lett.",
    volume = "87",
    pages = "142302",
    year = "2001"
}

@article{Cosyn:2018rdm,
    author = "Cosyn, W. and Pire, B.",
    title = "{Transversity generalized parton distributions for the deuteron}",
    eprint = "1806.01177",
    archivePrefix = "arXiv",
    primaryClass = "hep-ph",
    reportNumber = "CPHT-RR032.052018",
    doi = "10.1103/PhysRevD.98.074020",
    journal = "Phys. Rev. D",
    volume = "98",
    number = "7",
    pages = "074020",
    year = "2018"
}

@article{Cano:2003ju,
    author = "Cano, F. and Pire, B.",
    title = "{Deep electroproduction of photons and mesons on the deuteron}",
    eprint = "hep-ph/0307231",
    archivePrefix = "arXiv",
    doi = "10.1140/epja/i2003-10127-x",
    journal = "Eur. Phys. J. A",
    volume = "19",
    pages = "423--438",
    year = "2004"
}

@article{Kirchner:2003wt,
    author = "Kirchner, A. and Mueller, Dieter",
    title = "{Deeply virtual Compton scattering off nuclei}",
    eprint = "hep-ph/0302007",
    archivePrefix = "arXiv",
    doi = "10.1140/epjc/s2003-01415-x",
    journal = "Eur. Phys. J. C",
    volume = "32",
    pages = "347--375",
    year = "2003"
}

@article{Kovchegov:2012ga,
    author = "Kovchegov, Yuri V. and Sievert, Matthew D.",
    title = "{A New Mechanism for Generating a Single Transverse Spin Asymmetry}",
    eprint = "1201.5890",
    archivePrefix = "arXiv",
    primaryClass = "hep-ph",
    doi = "10.1103/PhysRevD.86.034028",
    journal = "Phys. Rev. D",
    volume = "86",
    pages = "034028",
    year = "2012",
    note = "[Erratum: Phys.Rev.D 86, 079906 (2012)]"
}

@article{Zhou:2013gsa,
    author = "Zhou, Jian",
    title = "{Transverse single spin asymmetries at small x and the anomalous magnetic moment}",
    eprint = "1308.5912",
    archivePrefix = "arXiv",
    primaryClass = "hep-ph",
    doi = "10.1103/PhysRevD.89.074050",
    journal = "Phys. Rev. D",
    volume = "89",
    number = "7",
    pages = "074050",
    year = "2014"
}

@article{Boer:2015pni,
    author = {Boer, Dani\"el and Echevarria, Miguel G. and Mulders, Piet and Zhou, Jian},
    title = "{Single spin asymmetries from a single Wilson loop}",
    eprint = "1511.03485",
    archivePrefix = "arXiv",
    primaryClass = "hep-ph",
    doi = "10.1103/PhysRevLett.116.122001",
    journal = "Phys. Rev. Lett.",
    volume = "116",
    number = "12",
    pages = "122001",
    year = "2016"
}

@article{Kovchegov:2021iyc,
    author = "Kovchegov, Yuri V. and Santiago, M. Gabriel",
    title = "{Quark sivers function at small $x$: spin-dependent odderon and the sub-eikonal evolution}",
    eprint = "2108.03667",
    archivePrefix = "arXiv",
    primaryClass = "hep-ph",
    doi = "10.1007/JHEP11(2021)200",
    journal = "JHEP",
    volume = "11",
    pages = "200",
    year = "2021",
    note = "[Erratum: JHEP 09, 186 (2022)]"
}

@article{Brodsky:2013oya,
    author = "Brodsky, Stanley J. and Hwang, Dae Sung and Kovchegov, Yuri V. and Schmidt, Ivan and Sievert, Matthew D.",
    title = "{Single-Spin Asymmetries in Semi-inclusive Deep Inelastic Scattering and Drell-Yan Processes}",
    eprint = "1304.5237",
    archivePrefix = "arXiv",
    primaryClass = "hep-ph",
    doi = "10.1103/PhysRevD.88.014032",
    journal = "Phys. Rev. D",
    volume = "88",
    number = "1",
    pages = "014032",
    year = "2013"
}

@article{Dumitru:2024pcv,
    author = {Dumitru, Adrian and M\"antysaari, Heikki and Tawabutr, Yossathorn},
    title = "{Polarized Dipole Scattering Amplitudes meet the Valence Quark Model}",
    eprint = "2407.08893",
    archivePrefix = "arXiv",
    primaryClass = "hep-ph",
    month = "7",
    year = "2024"
}

@article{Kovchegov:2018zeq,
    author = "Kovchegov, Yuri V. and Sievert, Matthew D.",
    title = "{Valence Quark Transversity at Small $x$}",
    eprint = "1808.10354",
    archivePrefix = "arXiv",
    primaryClass = "hep-ph",
    doi = "10.1103/PhysRevD.99.054033",
    journal = "Phys. Rev. D",
    volume = "99",
    number = "5",
    pages = "054033",
    year = "2019"
}

@article{Kovchegov:2013cva,
    author = "Kovchegov, Yuri V. and Sievert, Matthew D.",
    title = "{Sivers function in the quasiclassical approximation}",
    eprint = "1310.5028",
    archivePrefix = "arXiv",
    primaryClass = "hep-ph",
    doi = "10.1103/PhysRevD.89.054035",
    journal = "Phys. Rev. D",
    volume = "89",
    number = "5",
    pages = "054035",
    year = "2014"
}

@article{Kovchegov:2020kxg,
    author = "Kovchegov, Yuri V. and Santiago, M. Gabriel",
    title = "{Lensing mechanism meets small- $x$ physics: Single transverse spin asymmetry in $p^{\uparrow}+p$ and $p^{\uparrow}+A$ collisions}",
    eprint = "2003.12650",
    archivePrefix = "arXiv",
    primaryClass = "hep-ph",
    doi = "10.1103/PhysRevD.102.014022",
    journal = "Phys. Rev. D",
    volume = "102",
    number = "1",
    pages = "014022",
    year = "2020"
}

@article{Kovchegov:2022kyy,
    author = "Kovchegov, Yuri V. and Santiago, M. Gabriel",
    title = "{T-odd leading-twist quark TMDs at small x}",
    eprint = "2209.03538",
    archivePrefix = "arXiv",
    primaryClass = "hep-ph",
    doi = "10.1007/JHEP11(2022)098",
    journal = "JHEP",
    volume = "11",
    pages = "098",
    year = "2022"
}

@article{Cougoulic:2022gbk,
    author = "Cougoulic, Florian and Kovchegov, Yuri V. and Tarasov, Andrey and Tawabutr, Yossathorn",
    title = "{Quark and gluon helicity evolution at small x: revised and updated}",
    eprint = "2204.11898",
    archivePrefix = "arXiv",
    primaryClass = "hep-ph",
    doi = "10.1007/JHEP07(2022)095",
    journal = "JHEP",
    volume = "07",
    pages = "095",
    year = "2022"
}

@article{Hatta:2016wjz,
    author = "Hatta, Yoshitaka and Xiao, Bo-Wen and Yoshida, Shinsuke and Yuan, Feng",
    title = "{Single Spin Asymmetry in Forward $pA$ Collisions}",
    eprint = "1606.08640",
    archivePrefix = "arXiv",
    primaryClass = "hep-ph",
    reportNumber = "YITP-16-79",
    doi = "10.1103/PhysRevD.94.054013",
    journal = "Phys. Rev. D",
    volume = "94",
    number = "5",
    pages = "054013",
    year = "2016"
}

@article{Boussarie:2019icw,
    author = "Boussarie, Renaud and Hatta, Yoshitaka and Yuan, Feng",
    title = "{Proton Spin Structure at Small-$x$}",
    eprint = "1904.02693",
    archivePrefix = "arXiv",
    primaryClass = "hep-ph",
    doi = "10.1016/j.physletb.2019.134817",
    journal = "Phys. Lett. B",
    volume = "797",
    pages = "134817",
    year = "2019"
}

@article{Hagiwara:2020mqb,
    author = "Hagiwara, Yoshikazu and Hatta, Yoshitaka and Pasechnik, Roman and Zhou, Jian",
    title = "{Spin-dependent Pomeron and Odderon in elastic proton-proton scattering}",
    eprint = "2003.03680",
    archivePrefix = "arXiv",
    primaryClass = "hep-ph",
    doi = "10.1140/epjc/s10052-020-8007-6",
    journal = "Eur. Phys. J. C",
    volume = "80",
    number = "5",
    pages = "427",
    year = "2020"
}

@article{Mantysaari:2023xcu,
    author = {M\"antysaari, Heikki and Salazar, Farid and Schenke, Bj\"orn},
    title = "{Energy dependent nuclear suppression from gluon saturation in exclusive vector meson production}",
    eprint = "2312.04194",
    archivePrefix = "arXiv",
    primaryClass = "hep-ph",
    reportNumber = "INT-PUB-24-008",
    doi = "10.1103/PhysRevD.109.L071504",
    journal = "Phys. Rev. D",
    volume = "109",
    number = "7",
    pages = "L071504",
    year = "2024"
}

@incollection{Iancu:2003xm,
    author = "Iancu, Edmond and Venugopalan, Raju",
    editor = "Hwa, Rudolph C. and Wang, Xin-Nian",
    title = "{The Color glass condensate and high-energy scattering in QCD}",
    booktitle = "{Quark-gluon plasma 4}",
    eprint = "hep-ph/0303204",
    archivePrefix = "arXiv",
    doi = "10.1142/9789812795533_0005",
    pages = "249--363",
    year = "2003",
publisher="World Scientific"
}

@article{Morreale:2021pnn,
    author = "Morreale, Astrid and Salazar, Farid",
    title = "{Mining for Gluon Saturation at Colliders}",
    eprint = "2108.08254",
    archivePrefix = "arXiv",
    primaryClass = "hep-ph",
    reportNumber = "Universe 2021",
    doi = "10.3390/universe7080312",
    journal = "Universe",
    volume = "7",
    number = "8",
    pages = "312",
    year = "2021"
}

\end{document}